# Design rules for customizable optical materials based on nanocomposites

**DANIEL WERDEHAUSEN,**[1,2,3*] **ISABELLE STAUDE,**[2] **SVEN BURGER,**[4,5] **JÖRG PETSCHULAT,**[6] **TORALF SCHARF,**[3] **THOMAS PERTSCH,**[2,7] AND **MANUEL DECKER**[1]

[1]*Corporate Research & Technology, Carl Zeiss AG, Carl Zeiss Promenade 10, 07745 Jena, Germany*
[2]*Institute of Applied Physics, Abbe Center of Photonics, Friedrich Schiller University Jena, Albert-Einstein-Str. 15, 07745 Jena, Germany*
[3]*Nanophotonics and Metrology Laboratory, École Polytechnique Fédérale de Lausanne (EPFL), CH-1015 Lausanne, Switzerland*
[4]*JCMwave GmbH, Bolivarallee 22, 14050 Berlin, Germany*
[5]*Zuse Institute Berlin, Takustr. 7, 14195 Berlin, Germany*
[6]*Semiconductor Mask Solutions, Carl Zeiss SMT GmbH, Carl Zeiss Promenade 10, 07745 Jena, Germany*
[7]*Fraunhofer Institute for Applied Optics and Precision Engineering, Albert-Einstein-Str. 7, 07745 Jena, Germany*
*\*daniel.werdehausen@zeiss.com*

**Abstract:** Nanocomposites with tailored optical properties can provide a new degree of freedom for optical design. However, despite their potential these materials remain unused in bulk optical applications. Here we investigate the conditions under which they can be used for such applications using Mie theory, effective medium theories, and numerical simulations based on the finite element method. We show that due to scattering different effective medium regimes have to be distinguished, and that bulk materials can only be realized in a specific parameter range. Our analysis also enables us to quantify the range of validity of different effective medium theories, and identify design rules on how the material's free material parameters should be adjusted for specific applications.

## 1. Introduction

Nanocomposites have attracted a lot of interest, since their properties can be tuned by adjusting the concentrations, sizes, shapes, as well as constituent materials [1-4]. This, in principle, allows to design materials whose optical properties are tailored to the application at hand, which is especially promising for all applications that are limited to a small selection of materials, e.g. novel fabrication techniques like 3D printing [5, 6]. However, even though there has been a plethora of research, especially in the field of high-refractive-index-nanocomposites [1, 2, 4, 7, 8], these types of materials remain mostly unused in optical imaging systems whose thickness exceeds a few micrometers. This is partly due to challenges in the fabrication of such materials, but also a consequence of the fact that a detailed analysis and critical assessment of the theoretical requirements and limits has not yet been carried out. Furthermore, it remains unclear whether the resonances of high-refractive-index nanoparticles, the Mie resonances [9], can be used to tailor the effective refractive index of three dimensional bulk optical materials [10-12]. Finally, it is time for a thorough look at what constitutes a real effective optical material, and where the concept of an effective refractive index reaches its limits.

To fill these gaps, we here analyze the requirements for the use of nanocomposites as optical materials using Mie theory, effective medium theories, and finite element method (FEM) simulations. This allows us to derive simple design rules on how the particle size, shape, material, and concentration should be adjusted to achieve the desired effective optical properties (refractive index, scattering, and absorption). We moreover show that the fundamental connection between particle size and scattering necessitates the distinction of different effective medium regimes, and provide guidelines for the selection of a suitable effective medium theory in each of these regimes. This also allows us to elucidate the conditions under which the concept of an effective refractive index stands on solid ground. We then discuss two prototype nanocomposite materials (TiO$_2$ and Au nanoparticles), in order to identify the potential of nanocomposites for optical applications. We also clarify if the Mie-resonances can be utilized to tailor the effective refractive index and dispersion of a composite material. Our results reveal that it is possible to fabricate bulk optical nanocomposites in a limited parameter space, and bridge the gap between optical design and material fabrication.

## 2. The optical properties of composite materials

To describe the properties of composite materials that contain a large number of nanoparticles, we first discuss the response of a single particle. Afterwards, we show how effective medium theories can be used to treat systems that include a large number of identical particles.

*2.1 Multipole description for single particles*

In this paper, we mainly discuss spherical inclusions because fabrication techniques that enable an industrial-scale production of nanoparticles commonly yield close to spherical shapes. This also allows us to exploit that Mie theory [9] provides an analytical solution for the electromagnetic response of spheres of arbitrary size. However, our analysis can be readily extended to non-spherical particles [13]. Mie theory, in general, provides an analytical solution for the electromagnetic response of a sphere with permittivity $\epsilon_{\text{inc}}$ embedded in a host with permittivity $\epsilon_{\text{h}}$. The theory's main result is that the response of the sphere can be written as a multipole expansion, in which the different multipoles are driven by the corresponding multipole amplitudes of the incident wave. This allows determining the extinction ($\sigma_{\text{ext}}$), and scattering ($\sigma_{\text{scat}}$) cross sections [14]:

$$\sigma_{\text{ext}}^{\text{Mie}} = \frac{\lambda^2}{2\pi}\sum_{n=0}^{\infty}(2n+1)\text{Re}(a_n + b_n), \quad (1)$$

$$\sigma_{\text{scat}}^{\text{Mie}} = \frac{\lambda^2}{2\pi}\sum_{n=0}^{\infty}(2n+1)(|a_n|^2 + |b_n|^2), \quad (2)$$

where $a_n$ and $b_n$ are the electric and magnetic Mie coefficients, which correspond to the different multipoles (*n=1*: dipoles; *n=2*: quadrupoles). These coefficients, in turn, are functions of the relative refractive index $m = n_{\text{inc}}/n_{\text{h}} = \sqrt{\epsilon_{\text{inc}}}/\sqrt{\epsilon_{\text{h}}}$, and the size parameter $x = \pi n_{\text{h}} d_{\text{inc}}/\lambda$. Here $n_{\text{inc}}$ ($n_{\text{h}}$) corresponds to the refractive index of the inclusions (host), and $d_{\text{inc}}$ is diameter of the inclusions. Finally, the absorption cross section ($\sigma_{\text{abs}}$) follows from energy conservation as $\sigma_{\text{abs}} = \sigma_{\text{ext}} - \sigma_{\text{scat}}$. For dilute solutions, i.e. in the single scattering limit, this allows to directly determine the attenuation coefficient γ (as defined in $I(z) = I(0)\exp(-\gamma z)$, where $I(z)$ is the intensity) [14]:

$$\gamma_{Mie}^{ext} = \underbrace{N \cdot \sigma_{ext}^{Mie}}_{\text{Single scattering}} = \frac{8f}{\frac{4}{3}\pi d_{inc}^3} \cdot \sigma_{ext}^{Mie}, \quad (3)$$

where $N$ is the number density, and $f = 4/3\, N\pi(d_{inc}/2)^3$ is the volume fraction. To treat non-spherical particles the particles' multipole response simply has to be determined by other means, e.g. numerical simulations. Fig. 1(a) exemplarily depicts the attenuation coefficient obtained from equations (3) and (1) as a function of the wavelength and $d_{inc}/\lambda$ for $TiO_2$ spheres having a diameter of $d_{inc} = 140$ nm ($f = 0.1$ %). To highlight the multipolar character the contributions from the different multipoles up to the quadrupoles are also plotted individually. The figure illustrates that each component exhibits a distinct resonance, the so-called Mie resonance. These resonances are a characteristic feature of high-index-nanoparticles, and can be tailored by changing the particles' size, shape and material [15]. However, a power expansion of the Mie-coefficients yields that in the lowest order the electric dipole coefficient $a_1$ scales with $x^3$ ($x \propto d_{inc}/\lambda$), whereas both $b_1$ (magnetic dipole), and $a_2$ (electric quadrupole) scale with $x^5$ [14]. This shows that in the small particle limit only the electric dipole contributes to the particles' response. All second order corrections (second term in $a_1$, first terms in the expansion of $b_1$, and $a_2$) are of the order $x^5$ [14]. This is illustrated in Fig. 1(b), where the relative contributions of the different multipoles are depicted for the example from Fig. 1(a). The figure confirms that for small $d_{inc}/\lambda$ the response is solely determined by the electric dipole, and that the magnetic dipole and the two quadrupoles only play a role for larger values of $d_{inc}/\lambda$. This finding raises the question to what degree the different resonances can be used to tailor the optical properties of bulk materials. We emphasize, however, that all terms are also functions of the relative refractive index ($m$), which determines the relative magnitude of the higher order terms.

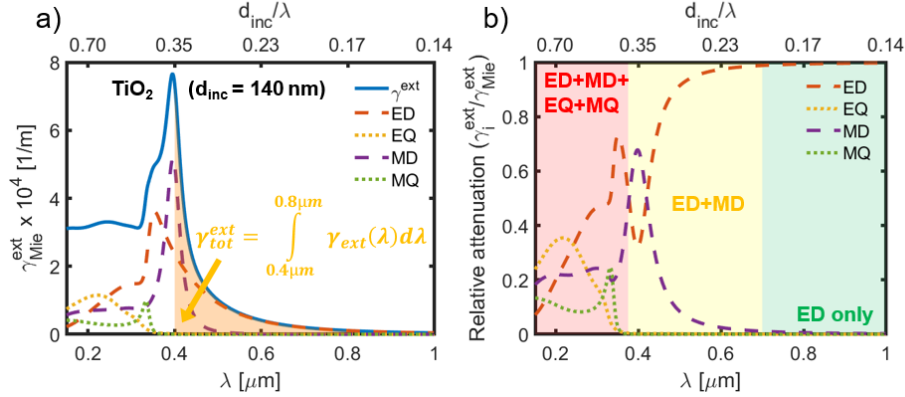

*Fig. 1: Attenuation coefficient as a function of the wavelength and $d_{inc}/\lambda$ for $TiO_2$ spheres at $d_{inc}$=140nm in a vacuum (a) and relative contribution of the different multipoles (b) for a dilute suspension at f=0.1% (ED, MD, EQ, MQ denote the electric (E) and magnetic (M) dipoles (D) and quadrupoles (Q), respectively). The shaded area in (a) corresponds to the "total attenuation in the visible" regime. The different colors in (b) highlight the regimes in which different multipole orders must be considered.*

*2.2 From single particles to an effective refractive index*

The refractive index is a powerful concept that allows reducing what microscopically is a complex multibody scattering problem to a macroscopic scalar measure. Generally, this concept is based on two approximations that limit its validity for composite mediums: Firstly, the building blocks of matter are assumed to be so small they can be treated as dipole scatterers, and secondly the medium is considered to be homogenous on the scale of the wavelength, so that the summation over the individual dipoles can be replaced by an integration [16, 17]. In fact, it can be shown that a wave propagating in a material consisting of such a continuous distribution of ideal dipoles is canceled out and replaced by a wave travelling at a lower speed [17, 18]. The refractive index then simply is the ratio of the propagation velocities of the new and the original wave. While these approximations are valid for conventional optical materials in the visible regime, they break down if higher order multipoles play a role, or the material is no longer homogenous on the scale of the wavelength. This discussion already shows that the optical properties of a composite material can only be described by an **effective refractive index** ($n_{\text{eff}}$) if the material fulfills both approximations to a large enough degree.

Theories that allow to calculate $n_{\text{eff}}$ from the properties of the constituents are called effective medium theories. The best known effective-medium theory is the **Clausius-Mossoti (CM)** equation:

$$\epsilon_{\text{eff}} = \epsilon_{\text{h}} \frac{1+\frac{16f}{d_{\text{inc}}^3}\alpha_{\text{inc}}}{1-\frac{8f}{d_{\text{inc}}^3}\alpha_{\text{inc}}}, \tag{4}$$

which connects the dipole polarizability of the inclusions ($\alpha_{\text{inc}}$) to the effective permittivity of the composite material. The effective refractive index of the system can be obtained from $n_{\text{eff}} = \sqrt{\epsilon_{\text{eff}}}$. However, it is important to be aware of key assumptions underlying the CM equation. These assumptions are that each particle is described by a dipole moment $\boldsymbol{p_{\text{inc}}} = \alpha_{\text{inc}}\boldsymbol{E_{\text{loc}}}$, and that the local field ($\boldsymbol{E_{\text{loc}}}$) matches the external field ($<\boldsymbol{E_{\text{loc}}}> = \boldsymbol{E_{\text{ext}}}$). The latter assumption is only true if the inclusions are randomly distributed, or are positioned on a primitive cubic lattice [19]. However, in these two cases the CM equation accounts for dipole-dipole interactions, i.e. multiple scattering, between the inclusions [20]. This is the reason why the CM equation remains accurate even for relatively large volume fractions ($f > 20\%$) [20]. But there are limits to the accuracy at high volume fractions, since impenetrable nanoparticles with a finite volume can never achieve a perfectly random distribution.

Starting from the CM equation the final task for inclusions of a finite size is to express the polarizability $\alpha_{\text{inc}}$ in terms of the constituents' properties, i.e. the size, shape and refractive index. Within Mie theory this can be achieved by dividing dipole component of the scattered field by the dipole amplitude of the incident wave, which yields [21]:

$$\alpha_{\text{inc}}^{\text{Mie}} = i\frac{3(d_{\text{inc}}/2)^3}{2x^3}a_1, \tag{5}$$

where $a_1$ is the first order Mie coefficient. By substituting equation (5) into equation (4) we arrive at the so-called **Maxwell-Garnett-Mie (MGM)** theory. Furthermore, if $a_1$ is replaced by the previously mentioned expansion, and all but the first order term are neglected one arrives at:

$$\alpha_{\text{inc}}^{\text{stat}} = (d_{\text{inc}}/2)^3 \frac{\epsilon_{\text{inc}}-\epsilon_{\text{h}}}{\epsilon_{\text{inc}}+2\epsilon_{\text{h}}}, \tag{6}$$

which is the expression that can also be obtained in the quasistatic limit, where the field is assumed to be constant across the particle. This equation is the basis of the original **Maxwell-Garnett (MG) theory**. The main difference between equations (5) and (6) is that the former takes retardation within the dipole term (i.e. the higher order terms in the expansion of $a_1$) into account, and is therefore valid up to larger particle sizes. As already discussed the next terms in the expansion of the Mie coefficients are the magnetic dipole and the electric quadrupole. If the magnetic dipole dominates over the electric quadrupole, the validity of the MGM theory can be extended to larger particle sizes by taking the contribution from $b_1$ into account. This leads to an "effective magnetic material", in that the material possesses a permeability $\mu_{\text{eff}}$ that differs from that of vacuum even if all components are non-magnetic [22]:

$$\mu_{\text{eff}} = \frac{x^3 + 3ifb_1}{x^3 - \frac{3}{2}ifb_1}. \tag{7}$$

The refractive index now readily follows from $n_{\text{eff}} = \sqrt{\epsilon_{\text{eff}}\mu_{\text{eff}}}$. In the following, we will refer to this expression for $n_{\text{eff}}$ as the **MGMm theory**, to highlight that includes the magnetic dipole response. Microscopically, the magnetic dipole radiation is caused by excitation of strong circular displacement currents inside the sphere [15, 23]. But we stress again that in the lowest order both $b_1$ and $a_2$ scale with the same power of $x$, and it is therefore not in general true, that taking the magnetic dipole contribution into account allows to extend the validity of the MGM theory. This is shown later, on the comparison of metallic and dielectric nanospheres.

Within the effective medium theories, the attenuation coefficient can be directly obtained from the imaginary part of the effective refractive index:

$$\gamma_{\text{eff}}^{\text{ext}} = \frac{4\pi}{\lambda} \cdot \text{Im}(n_{\text{eff}}). \tag{8}$$

As opposed to equation (3), which only holds true in the single scattering limit this expression accounts for multiple scattering to a certain degree (see discussion of equation (4)).

Finally, it is important to mention that both expressions for the polarizability are only valid, if the optical properties of the inclusions can be described by a permittivity. This limits the validity of the theories for small particles, since the permittivity is a macroscopic quantity that is only meaningful if the inclusions contain a sufficiently high number of atoms [24]. For particles that are very small, yet contain a large enough number of particles, it is moreover important to keep in mind that the permittivity becomes size depended due to confinement effects and surface scattering [25-27]. In the case of metallic nanoparticles this can be accounted for by replacing the bulk damping rate by a size dependent one, which accounts for the increased amount of surface scattering [26]. Throughout this paper we include this relationship in our modelling of metallic nanoparticles. In contrast, for dielectric particles it is well known that the nanoparticles' optical properties become size dependent due to quantum confinement effects at diameters in the order of the exciton Bohr radius, which highly depends on the choice of material [25].

## 3. Effective medium regimes

The above discussion illustrates that the macroscopic optical properties of a nanocomposite depend heavily on the particle size, and that there are fundamental limits to the validity of all effective medium theories. As discussed, the main difference between the original MG theory and the MGM theory is that the latter takes retardation within the dipole term into account. This has fundamental consequences that can be best illustrated by expanding the CM equation up to the first order of $N\alpha$:

$$n_{\text{eff}} = \sqrt{\epsilon_{\text{eff}}} \approx \underbrace{n_{\text{h}}\big(1 + 2\pi N \text{Re}(\alpha_{\text{inc}})\big)}_{\text{Re}(n_{\text{eff}})} + \text{i} \cdot \underbrace{n_{\text{h}} 2\pi N \text{Im}(\alpha_{\text{inc}})}_{\text{Im}(n_{\text{eff}})} + O((n\alpha)^2), \tag{9}$$

where we assumed the host to be lossless ($\text{Im}(n_{\text{h}}) = 0$), and the permeability of the composite medium to be unity. This relationship shows that for dilute solutions the real part of $\alpha$ only affects the real part of $n_{\text{eff}}$, whereas its imaginary part is determined by the imaginary part of $\alpha$. The difference between the different effective medium theories can now be understood by analyzing the respective expressions for $\alpha$ (equations (5) and (6)): The quasistatic expression (equation (6)), which is used in the MG theory, only yields a non-zero imaginary part for $\alpha$ if the inclusions are absorptive, whereas the dipole polarizability from Mie theory (equation (5)) also yields a non-zero imaginary part if the particles have non-zero scattering cross section [14]. This shows that the MGM and MGMm theories account for scattering losses, and can consequently predict losses, i.e. a non-zero imaginary part of the refractive index, even if two non-absorbing materials are used [28]. As discussed below, this implies that, in the presence of scattering, the effective refractive index from these theories no longer has the same validity as the refractive index of a homogenous material.

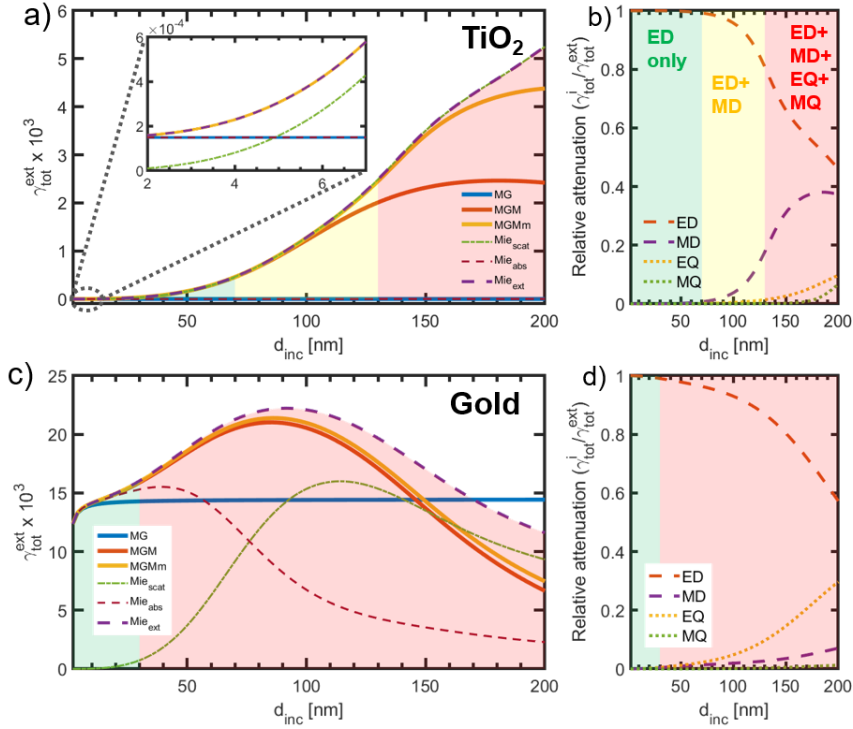

*Fig. 2: Total (spectrally integrated) attenuation in the visible regime for TiO$_2$ (a) and gold (c) nanoparticles at f=0.1% in PMMA as a function of the particle size. Mie$_{ext}$, Mie$_{scat}$, and Mie$_{abs}$ denote the total attenuation, scattering, and absorption losses as obtained from Mie theory, respectively. In (b) and (d) the relative weight of the multipoles ($\gamma_{tot}^{i}/\gamma_{tot}^{ext}$) are shown. Green area: response solely defined by electric dipole; yellow area: magnetic dipole plays a role; red area: At least one of the quadrupoles has to be considered. The inset of (a) illustrates that for TiO$_2$ scattering plays a role for sizes over 2 nm, and dominates over absorption for sizes larger than 5 nm.*

*3.1 Attenuation in composite materials*

The fundamental connection between retardation and density fluctuations in the microscopic, and incoherent scattering in the macroscopic picture, inherently limits the validity of any effective refractive index. In the following we will analyze at what particle sizes the different effects start to play a role, and discuss the consequences for the validity of the different effective medium theories. We will perform the analysis on the example of two prototype systems: Firstly, the previously used example of TiO$_2$-spheres, and secondly, a nanocomposite containing gold-nanospheres. In both cases we use PMMA as a widely used optical polymer as the host material. We selected TiO$_2$ because of its high-refractive index ($n > 2.4$) and good transparency in the visible regime, which make it perfectly suited to investigate the properties of high-refractive-index nanocomposites [1, 4, 15]. Gold nanoparticles were chosen for comparison, because their dominant plasmon resonance leads to drastically different optical properties. Since the focus of this paper is on optical materials, we limit our analysis to the visible spectral range, but it can be readily extended to other regimes. To be able to quantify the response throughout the respective wavelength range we use the "total attenuation in the visible regime" (orange area in Fig. 1(a)):

$$\gamma_{\text{tot}}^{i} = \int_{400\,nm}^{800\,nm} \gamma_{i}^{\text{ext}}(\lambda) d\lambda, \qquad (10)$$

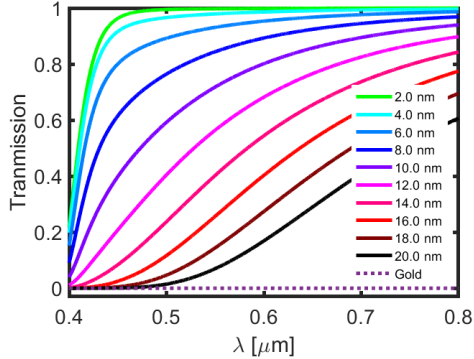

*Fig. 3: Transmission of a coherent beam of light through a 1 mm thick slab of a TiO$_2$-nanocomposite in PMMA for different particle sizes and f=12.5% (obtained from $I(d)/I_0 = exp(-\gamma z)$) and the MGMm theory). The transmissivity rapidly decreases as the size is increased due to scattering. The dotted line illustrates, that for a corresponding gold-nanocomposite at f=2.5%, the transmission is zero at all particle sizes because of the large amount of absorption.*

where $\gamma_i^{ext}(\lambda)$ is taken from either equation (8) or equation (3). To allow the use equation (3), which follows from Mie theory in the single scattering limit and is therefore only valid for dilute suspensions, as the ground truth, we investigate the properties of both prototype systems at relatively low volume fractions (f=0.1%). As already discussed we, furthermore, account for the size dependence of the gold nanoparticles' permittivity due to surface scattering through a size-dependent damping rate, whereas we assume the permittivity of TiO$_2$ to be size-independent for this analysis.

Fig. 2(a) depicts $\gamma_{tot}$, obtained from both Mie theory, as well as the different effective medium theories, as a function of the particle size for the TiO$_2$ prototype system. We also used Mie theory to disentangle the contributions from scattering and absorption. The figure's inset illustrates that scattering causes the total attenuation to increase for particle sizes larger than 2 nm, and eventually dominates over absorption for $d_{inc} > 5$ nm. Since the original MG theory can't account for scattering losses, this consequently already is where it loses its validity. For larger particles both, the MGM and the MGMm theories correctly predict the amount of attenuation for sizes up to approximately 70 nm, where the MGM theory starts to underestimate the losses. Fig. 2(b) illustrates that this diameter range ($d_{inc} < 70$ nm) is the regime, where only the electric dipole ($a_1$) has to be considered (green area). However, for $d_{inc} > 70$ nm the magnetic dipole ($b_1$) starts to play a role, which is why only the MGMm theory can correctly predict the attenuation in this regime (yellow area). Finally, for $d_{inc} > 130$ nm (red area), none of the effective medium theories can correctly predict the total attenuation (Fig. 2(a)), since the electric quadrupole component starts to emerge.

Figs. 2(c) and 2(d) show that the system containing gold nanoparticles exhibits a significantly different behavior: Fig 2(c) illustrates that due to the high amount of absorption, scattering only contributes significantly for $d_{inc} > 30$ nm. As a result, the original MG theory provides a good estimate of the total attenuation up to much larger particle sizes than for the first system. Note, first of all, that this is only due to the very high amount of absorption. In absolute numbers, the total amount of scattering is larger than for the TiO$_2$ nanocomposite. Secondly, the increase of the absorption at very small radii is caused by the size dependence of the permittivity, and is therefore correctly reproduced by all theories. Finally, the decrease of $\gamma_{tot}^{ext}$ for $d_{inc} > 100$nm (Fig. 6(a)) is simply a result of the redshift of a part of the plasmon resonance out of the visible spectral range with increasing particle size. Furthermore, Fig. 2(c) also shows that even though the MGMm theory provides a slightly better approximation for larger radii, the MGM and the MGMm theory start to underestimate the total attenuation at roughly the same point. This is because the electric quadrupole dominates over the magnetic dipole at all radii (see Fig. 2(d)). This illustrates that, as previously mentioned, the relative refractive index *m* determines which of the higher order terms is dominant.

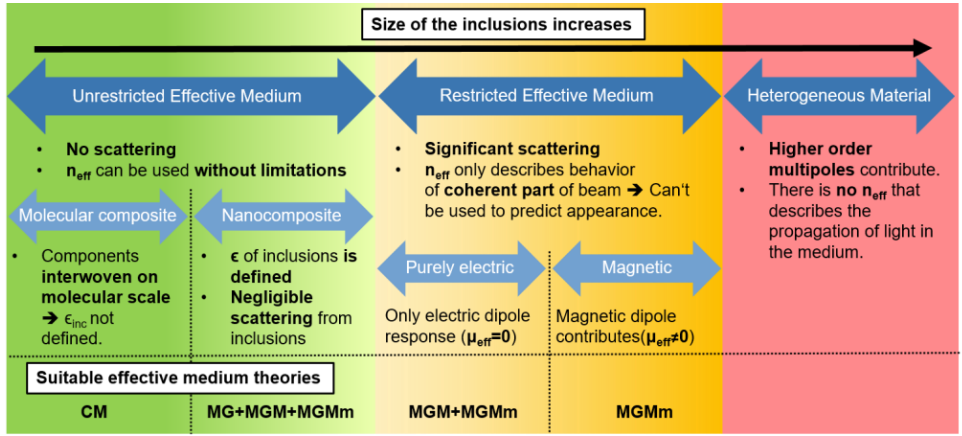

*Fig. 4: Effective medium regimes, and suitable effective medium theories (MG – Maxwell-Garnett; MGM – Maxwell-Garnett-Mie; MGMm Maxwell-Garnett-Mie with magnetic dipole response). Note that the "magnetic" regime within the "restricted effective medium" regime only exists if the magnetic dipole dominates over the electric quadrupole.*

To demonstrate how drastically the optical properties, depend on the particle size, and differ between the two prototype systems, Fig. 3 presents the transmission of a coherent beam of light through a 1mm thick slab of both systems as a function of the wavelength at a volume fraction of $f = 12\%$ ($TiO_2$) and $f = 2.5\%$ (gold). For both systems we determined the respective volume fraction from the condition that we required the change of the refractive index at the *d*-line ($\lambda_d = 587.56$ nm) relative to the host to be $\Delta n_d = 0.1$. The figure confirms that for the $TiO_2$-nanocomposite scattering causes the transmittivity to rapidly decrease for particle diameters exceeding 4 nm. Note, that the rapid drop of the transmission at short wavelengths is caused by the onset of the intrinsic absorption of $TiO_2$. Moreover, the dotted line in Fig. 3 demonstrates that due to the high amount of absorption the gold-nanocomposite does not transmit any light.

Finally, it is important to point out that through a periodic arrangement of the structures it is possible to design metamaterials, in which no incoherent scattering takes place even if the size or unit cell of the structures is in the order of the wavelength [11, 12, 29, 30]. Such a material is called a photonic crystal [31]. In fact, in a perfect photonic crystal scattering does not play a role, since, just like electrons in a crystal, the photons in such a material propagate as Bloch waves. The optical properties of a photonic crystal are therefore determined by the coupling of the plane waves that propagate outside the photonic crystal to the Bloch waves within. However, it has been shown that in a specific regime the refraction properties of such a material, that is the bending of the rays upon going through the interface, can also be described by an effective refractive index [29]. In fact, this effective refractive index can be readily determined from the photonic band structure [29]. However, it is important to keep in mind that such a material is not an effective material in the sense that we have been discussing so far (Fig. 4). The reason for this is that the refractive index of a conventional material follows from a microscopic averaging procedure, and therefore not only provides information about the bending of light at an interface, but also about the macroscopic fields within the material. In contrast, the effective refractive index determined from the band structure of a photonic crystal merely describes the bending of a ray according to Snell's law. This fundamental difference also has consequences for a phenomenological black-box-like description of a photonic crystal [30, 32]: The effective refractive index that describes the direction change is not the necessarily the same index that has to be used in the Fresnel formulas. It consequently does not always describe the changes in intensity, phase, and polarization of the different components [30]. Furthermore, the effective refractive index can also depend on the angle of incidence [29, 30].

*3.2 From homogenous to heterogeneous materials*

The discussion of Fig. 3 also reveals a fundamental limitation of all composite mediums: For a "homogenous" medium in the absence of fluctuations, ray optics can be used to predict the appearance of the material to an observer at any point in space. This, in contrast, no longer holds true for mediums in which a significant amount of scattering occurs. In such a medium, the effective refractive index obtained from the MGM or the MGMm theory only provides information about the attenuation of a coherent beam of light. It can consequently only be used to determine the appearance of the material to an observer who is looking straight into the beam through an infinitely small aperture. However, a slab of material that is predicted to have a low transmittivity might overall appear opaque (scattering dominates over absorption) or simply black (absorption dominates over scattering). The macroscopic appearance of a composite material therefore depends heavily on the particle size, which necessitates the distinction of different effective regimes (Fig. 4). To visualize the drastic differences between these regimes, we also performed FEM simulations on the propagation of a Gaussian beam through the prototype $TiO_2$-nanocomposite at different particle sizes using the commercially available software JCMsuite (Fig. 5). For these simulations, we kept the volume (area) fraction constant at $f = 40\ \%$, and varied only the particle size to ensure comparability. We chose four different particle sizes ($d_{\text{inc}} = 0$ nm, 6 nm 100 nm, and 200 nm), which are characteristic for the four effective medium regimes. For each particle size we then placed the corresponding number of particles randomly but without overlap into the area. All simulations were performed at a wavelength of 800 nm.

The left panels in Figs. 4 and 5 illustrate that if the inclusions in a composite material don't exceed atomic/molecular scale, and the different components in the material are well interwoven on the molecular scale, the medium behaves like a conventional **homogenous** material. The FEM simulation in the left panel of Fig. 5 shows that the shape of a beam of light propagating through such a **molecular composite** remains completely intact. The medium consequently is an **unrestricted effective medium**, in that scattering doesn't play a role and the concept of an (effective) refractive index is fully valid. This approach of doping a material with additional atoms or molecules to change the refractive index is used both in glasses [33] and polymers [8]. However, since the optical properties of the molecular inclusions can't be described by a macroscopic permittivity (sec. 2.2), the MG-type effective medium theories can't be used to determine the material's refractive index. Instead, the CM equation and the molecular polarizabilities have to be used. In contrast, the **nanocomposite sub-regime** of the **unrestricted effective medium regime** (right part of the green panel in Figs. 4 and 5) is reached for mediums that contain distinct particles, whose optical properties can be described by a macroscopic permittivity, but are so small that scattering doesn't play a role. In this regime, the concept of an effective refractive index is consequently still valid without limitations. This is directly visualized by the FEM simulation in the green panel of Fig. 5, which visualizes that in such a material the shape of a gaussian beam remains unaffected by scattering despite the high number of nanoparticles in the material. For this simulation we used a particle diameter of 6 nm, which corresponds to a ratio of $d_{\text{inc}}/\lambda = 0.0075$. Both the original and the extended Maxwell-Garnett effective medium theories can therefore be used to determine $n_{\text{eff}}$ for such a material. However, as the particle size is further increased, scattering starts to play a significant role, and the materials enter the **restricted effective medium regime** (yellow panel). In this regime absorption isn't the only mechanism that attenuates a coherent beam of light anymore, and $n_{\text{eff}}$ consequently provides no information about the origin of the losses, i.e. the incoherent part of the beam. This is confirmed by the FEM simulations depicted in Fig. 5 (yellow panel) using a particle size of 100 nm ($d_{\text{inc}}/\lambda = 0.125$). The figure illustrates that in this regime the beam retains some of its directionality but is quickly separated into different bundles. This shows that scattering significantly contributes to the overall attenuation of a coherent beam of light in such a material. Furthermore, if the magnetic dipole highly dominates over the electric quadrupole in a certain radius range, this regime also contains the **magnetic regime**, which is distinguished by non-unity permeability even if only nonmagnetic materials are used. Finally, at even larger diameters higher order multipoles start to contribute to the particles' response. In this regime, even if the nanoparticles still formed a continuous distribution, the presence of the electric quadrupole would require the electric displacement to be written as $\boldsymbol{D}(\omega,\boldsymbol{r}) = \epsilon_0 \boldsymbol{E}(\omega,\boldsymbol{r}) + \boldsymbol{P}(\omega,\boldsymbol{r}) - \frac{1}{2}\nabla \boldsymbol{Q}(\omega,\boldsymbol{r})$ [24], and the optical properties of the medium could therefore no longer be described by a conventional (effective) refractive index. Furthermore, as we have shown in the previous section, real nanoparticles with a non-zero quadruple component are too big in relationship to the wavelength ($d_{\text{inc}} = 130$ nm for TiO$_2$) for the particles to be able to still form a continuous distribution. Such a material is therefore located in the **heterogeneous regime** (red panel in Fig. 4), and can no longer be accurately described by any effective medium theory. In fact, a nanocomposite in this regime scatters light so effectively that a beam of light loses its directionality after very short propagation distances. This is confirmed by the FEM simulations depicted in the red panel of Fig. 5 for a particle size of 200 nm ($d_{\text{inc}}/\lambda = 0.25$). Finally, it is important to mention that the homogeneity requirement is also violated if the blending of the nanoparticles into the host is not perfect. Therefore, it is possible that a composite material is located in the unrestricted or even heterogenous regime, even if the inclusions themselves are small enough to avoid scattering.

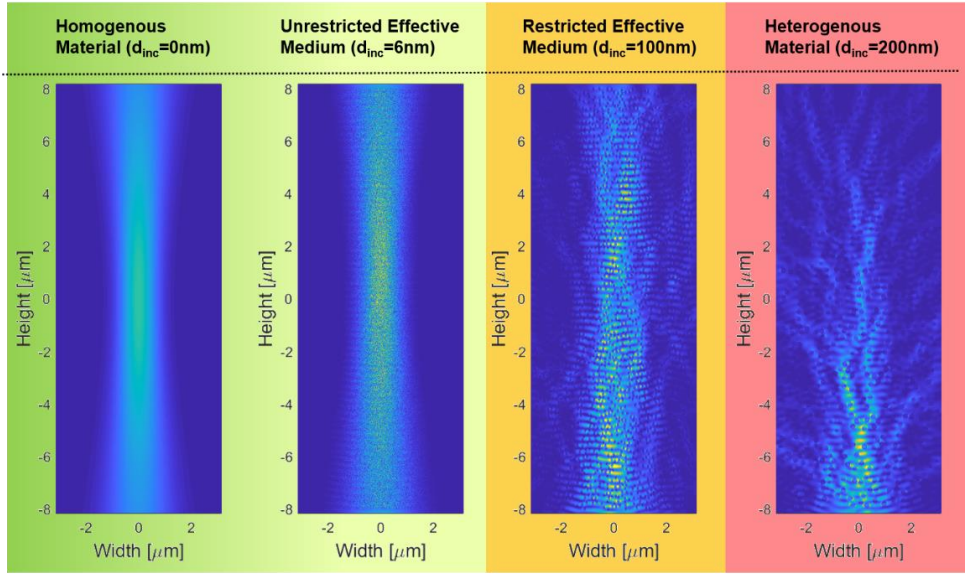

*Fig. 5: 2D FEM simulations visualizing the propagation of a Gaussian beam (λ=800nm) in the effective medium regimes at an area fraction of f =40% for TiO$_2$ in PMMA. The beam can propagate unaffected through an "unrestricted effective" medium, but is quickly separated into different bundles in a "restricted" medium. In a heterogeneous material the directionality is completely lost after several wavelengths.*

In summary, we have discussed that an effective refractive index can only be used without restrictions in the unrestricted refractive medium regime (Fig. 4). Outside of this regime, fundamental restrictions to the applicability of the effective refractive index arise. However, the existence of the unrestricted effective medium regime also shows that it is indeed possible to design bulk optical nanocomposites. Building on this, we dedicate the next section to the potential of nanocomposites as tailored optical materials in optical system design.

## 4. Design rules for nanocomposites

A conventional optical design software based on ray optics can be used to determine the optical properties that are optimal for a material in a specific optical system. The parameters that are commonly used by optical designers for this purpose are: The refractive index $n_d$ ($\lambda_d = 587{,}56$ nm), and the Abbe number ($\nu_d$) as a measure for the total amount of dispersion in the visible regime ( $\nu_d = n_d - 1/(n_F - n_C)$, with $\lambda_F = 486{,}13$ nm and $\lambda_C = 656{,}27$ nm). Furthermore, the overall design of the optical system also provides insight as to how much attenuation and scattering can be tolerated. However, in order to determine how the free parameters of the composite material have to be tailored to provide a material with the required properties, a suitable effective medium theory is indispensable. In this final section, we elaborate on how these parameters have to be adjusted to achieve the desired result.

Firstly, we analyze whether the characteristic Mie-resonances (Fig. 1) can be exploited to tailor $n_d$ or $\nu_d$, since this would provide a powerful degree of freedom for the design of new materials. To answer this question Fig. 6(a) depicts both quantities as a function of the particle size at f = 12%. We here again use the volume fraction, which fulfills $\Delta n_d = 0.1$. The figure demonstrates that for $d_{inc} < 10$ nm $n_d$ or $\nu_d$ hardly depend on the particle size. This is because at these diameters the resonances are located at very short wavelengths, and therefore don't affect the behavior in the visible spectral range. The refractive index is consequently solely defined by the volume fraction. However, Fig. 6(a) demonstrates that for $d_{inc} > 10$ nm the Abbe number changes significantly as the particle size increases, whereas $n_d$ only exhibits a weak size dependence. This illustrates that in this regime first the electric then also the magnetic dipole resonances start to affect the material's dispersion, since the respective resonances shift closer to the visible wavelength range. We emphasize, however, that this size range is located in the restricted effective medium regime, where a significant amount of scattering occurs (Fig. 4). Such a material is therefore not suited for bulk applications. For applications for which bulk nanocomposites are required, it is essential to choose a particle size that is located in the unrestricted effective medium regime. However, for thin film applications, e.g. if it is necessary to match dispersion of two materials over a broad wavelength range, the Mie resonances can be used as an additional degree of freedom to tailor the optical properties.

The second design parameter for nanocomposites is the volume fraction. Fig. 6(b) demonstrates that varying the volume fraction allows to tune both $n_d$ or $\nu_d$ along a trajectory that is defined by the constituent materials. This illustrates that the selection of the constituents, and the volume fraction are critical to reach the required values for $n_d$ or $\nu_d$. Furthermore, the discussion of Fig. 3 also showed that in addition to defining the behavior of the refractive index, the material choice is also essential for keeping the overall attenuation as low as possible.

At this point we also want to elaborate on the concentrations that were necessary to achieve $\Delta n_d = 0.1$ (TiO$_2$-nanocomposite: $f = 12\%$ and gold-nanocomposite: $f = 2.5\%$ ). These volume fractions are supposed to represent the minimum concentrations, which achieve a significant impact in optical systems (increased performance and/or reduced size). For the TiO$_2$-nanocomposite at $d_{inc} = 4$ nm the required concentration of $f = 12\%$ corresponds to 726000 nanoparticles per volume $\lambda_d^3$. In comparison, gold nanoparticles have a much higher polarizability due to their resonant character, but the concentration of $f = 2.5\%$ still corresponds to 151000 particles per $\lambda_d^3$. This leads us to a fundamental design rule: In order to achieve a significant change of the real part of the refractive index, a high number of nanoparticles per volume $\lambda^3$ are required. This can pose major challenges when synthesizing a homogeneous nanocomposite, since the particles tend to agglomerate [34]. Note again that the actual number depends on the particles' polarizability, which in turn depends on the particle size. If there are not enough particles per volume $\lambda^3$, the composite materials' response will be dominated by the atomic/molecular scatters that make up the host. The inclusions can, however, still significantly change the imaginary part of the refractive index, if they are highly absorptive or scatter light effectively. The necessity for high volume fractions in the order of 10% also explains why intrinsic absorption has to be completely eliminated.

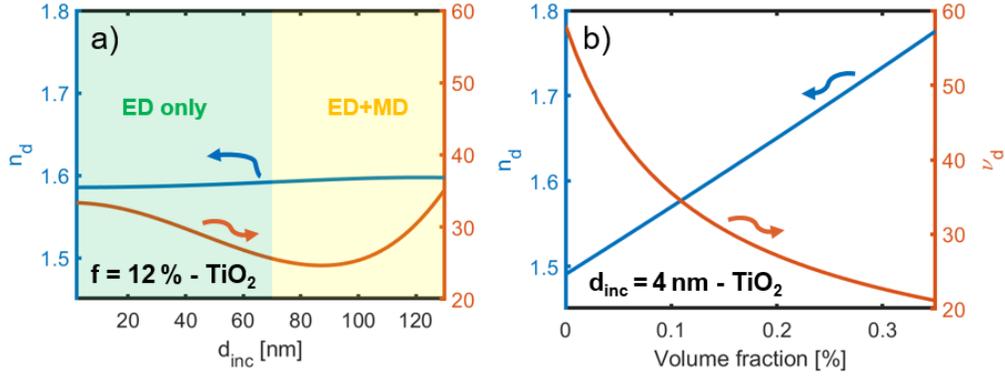

*Fig. 6: (a) $n_d$ and $v_d$ as a function of the particle size for TiO$_2$ nanoparticles in PMMA at f=12%: $n_d$ depends only weakly on the particle size, whereas the tail of the electric dipole resonance starts to affect $v_d$ for $d_{inc}>10nm$. The magnetic dipole contributes for $d_{inc}>70nm$. (b) $n_d$ and $v_d$ as a function of the volume fraction at $d_{inc}=4nm$. Adjusting the volume fraction changes the optical properties along a trajectory defined by the constituent materials.*

Finally the shape is an additional parameter that can also be used to tailor the nanoparticles' electromagnetic response [15, 35]. Shapes other than spheres can be readily incorporated into our framework by calculating the nanoparticles' multipole response and their respective polarizabilities by other means (e.g. FEM simulations [36]). Using shapes without full rotational symmetry would allow to design materials with a customized birefringence. For this purpose the inclusions have to be aligned in a defined direction. This is particularly interesting for highly nonlinear nanocomposites [37], since it allows to use the material's degrees of freedom to fulfill the phase matching condition in bulk materials.

## 5. Summary & Conclusion

Optical applications, especially imaging applications require homogenous materials in which scattering is reduced to a minimal level. In this paper, we have shown that there is a regime in which nanocomposites can fulfill these requirements. This regime, which we call the "unrestricted effective medium regime", is reached for particle sizes (significantly) below 4 nm. This demonstrates that the particle sizes used in nanocomposites until now have been mostly too large for bulk optical applications. Furthermore, the onset of scattering for larger particles necessitates the distinction of different effective medium regimes between "homogenous" and "heterogeneous" materials (Fig. 4). This distinction emphasizes that the effective refractive index can only be used without limitations in the "unrestricted effective medium regime", whereas fundamental limitations arise outside of this regime. Our analysis has also demonstrated that for dielectric particles, there exists a regime within the "restricted effective medium regime" with a non-unity permeability even if all components are non-magnetic. We have also emphasized what effective medium theory is valid in each regime and highlighted their limitations. Finally, we have discussed how the free parameters of a nanocomposite (the constituent materials, the concentrations, particle sizes and shapes) should be adjusted to achieve a material with the desired optical properties both in terms of attenuation as well as dispersion ($n_d$ and $v_d$). This bridges the gap between an optical design software, which can be used to optimize the optical properties of a material, and the design and fabrication of the material itself.

## 6. Funding

This project has received funding from the European Union's Horizon 2020 research and innovation programme under the Marie Sklodowska-Curie grant agreement No. 675745.